# A Security Framework for Ethereum Smart Contracts

Antonio López Vivar[a], Ana Lucila Sandoval Orozco[a], Luis Javier García Villalba[a,*]

[a]*Group of Analysis, Security and Systems (GASS)*
*Department of Software Engineering and Artificial Intelligence (DISIA)*
*Faculty of Computer Science and Engineering, Office 431*
*Universidad Complutense de Madrid (UCM)*
*Calle Profesor José García Santesmases 9, Ciudad Universitaria, 28040 Madrid, Spain*

**Abstract**

The use of blockchain and smart contracts have not stopped growing in recent years. Like all software that begins to expand its use, it is also beginning to be targeted by hackers who will try to exploit vulnerabilities in both the underlying technology and the smart contract code itself. While many tools already exist for analysing vulnerabilities in smart contracts, the heterogeneity and variety of approaches and differences in providing the analysis data makes the learning curve for the smart contract developer steep. In this article the authors present ESAF (Ethereum Security Analysis Framework), a framework for analysis of smart contracts that aims to unify and facilitate the task of analysing smart contract vulnerabilities which can be used as a persistent security monitoring tool for a set of target contracts as well as a classic vulnerability analysis tool among other uses.

## 1. Introduction

Blockchain and smart contracts [1] are becoming more and more popular partly due to the expansion related to Bitcoin [2] and also because of all their potential applications, such as: identity management [3] [4], electronic voting [5] [6], banking

*Corresponding author
Email addresses:* alopezvivar@fdi.ucm.es (Antonio López Vivar),
asandoval@fdi.ucm.es (Ana Lucila Sandoval Orozco), javiergv@fdi.ucm.es (Luis Javier García Villalba)

and financial services [7], supply chain [8], IoT [9], online gaming [10] [11], digital coupons [12] and medical information [13] for example. Although they are usually associated with the Ethereum platform [14], today there are many platforms that make use of them ([15] presents an updated list of all existing smart contract platforms). In this article, we will focus on Ethereum smart contracts, although much of the security concepts can be extended to other platforms.

Smart contracts are software programs that are executed in a decentralised manner based on blockchain technology. Like all computer programs, they are susceptible to vulnerabilities in their code, and like all emerging technologies, they are the focus of hackers who will try to exploit these vulnerabilities. To date, a wide variety of vulnerability analysis tools for smart contracts exists with different approaches, algorithms, input and output formats of results, making difficult for the developer of smart contracts to take advantage of them. After an exhaustive compilation of most of the vulnerability analysis tools in smart contracts, authors found the variety of existing tools, seen individually, means a relatively high learning curve for smart contract developers. When a smart contract developer first faces the task of analyzing his code he finds that there are a multitude of tools and solutions, each with different installation requirements, required dependencies, etc. In addition, the operation of these tools, although similar (many are based on symbolic execution), does not produce the same results, nor do all the tools analyze or detect the same vulnerabilities. Some are more effective than others in searching for various types of vulnerabilities. This is why the learning curve of all these tools is high.

In this paper, the authors present a framework that brings together all the combined power of the existing tools, with an interface that is as simple as possible and an output format of the analysis data that combines as much information as possible (without overlapping between the tools). Developers can carry out vulnerability analysis of their smart contracts by leveraging the strengths of each existing tool without having to worry about implementing and keeping all of them updated, and to provide a standardised output model that avoids "noise".

The main contributions of this work are, on the one hand, to study, organize and test the different existing security tools in smart contracts, showing their characteristics, dependencies, installation requirements, etc. and, on the other hand, the development of a framework with several tools that facilitate the analysis of vulnerabilities in smart contracts using the different tools analyzed in a combined way. The idea behind this framework is to make it easier for developers of smart contracts to analyze vulnerabilities in their smart contracts without having to worry about the different installation and operating requirements of each of these tools. Likewise, the developed framework has the potential to be used in persistent security analysis tasks of the blockchain by "monitoring" certain smart contracts of our interest throughout their useful life that even if they were not vulnerable at the beginning, later on with the update of the detection algorithms or the incorporation of new tools, they may turn out to have vulnerabilities.

The remaining paper is structured as follows: Section 1 is a brief introduction. Section 2 presents the different families of vulnerabilities that smart contracts can present, in addition to the tests carried out with the different tools currently available for analyzing vulnerabilities in smart contracts. In Section 3 The ESAF framework is presented, making a description of its architecture, implementation, and operation. Finally, the Section 5 presents the conclusions and future work of this article.

## 2. Background And Related Work

While blockchain is designed and supported by widely studied and tested cryptographic algorithms, smart contracts, like any piece of software, are likely to contain security vulnerabilities in their source code. Given the immutable nature of the blockchain, coupled with the fact that it can operate in sensitive domains such as financial or healthcare, it poses a serious security threat, as it has in the past. In addition to the flaws in the smart contract code, there are the inherent vulnerabilities in the blockchain and the vulnerabilities in the EVM (Ethereum Virtual Machine [16]). The following is a list of the most important families of vulnerabilities that Smart Contracts are exposed to. Some of them have been mitigated by Solidity compiler updates, although it is not always possible to remove them completely. Table 1 presents a summary of the types of vulnerabilities that occur in smart contracts. (The interested reader is able to find more information about these vulnerabilities in the works of [17] [18]).

Nowadays there are many tools for the analysis of vulnerabilities in smart contracts and new ones appearing all the time. We have focused in this work on analyzing vulnerabilities in the source code of smart contracts, although there are other works that focus on vulnerabilities in consensus protocols, such as [19], [20] or formal verification tools like [21].

Smart contracts can be written in several high-level languages, being the most used language Solidity. When the smart contract is compiled, the high-level language is translated into a bytecode that will execute the EVM (and that is what is stored in the blockchain). Most of the tools support the analysis of smart contracts in bytecode, although some also allow to analyze the source code written in Solidity.

Regarding the analysis methods used by the existing tools, they can be included in the following ones:

- **Static analysis**: static software analysis [22] is a way of studying the behaviour of a program from its compiled binary code without executing it, looking for known patterns that often lead to vulnerabilities.

- **Dynamic analysis**: dynamic software analysis [23] acts in the execution phase of the program, detecting vulnerabilities that could have gone unnoticed during the static analysis (for example due to the existence of obfuscated code or resistant packaging) at the same time that it serves to verify those found by this one.

- **Formal verification**: through this type of analysis [24] a verification is carried out using formal mathematical methods to test the specific properties of the code using theorems provers like SMT Z3 and Coq among others.

On the other hand, after trying and running all the above tools, we have found both useful tools and tools that either because they are complex to install or because they do not add value.

It is important to make a clear differentiation between those that use bytecode and Solidity since they focus on different types. For example, those that only use Solidity are more focused on the creation of graphs, inclusion of new code, and locating errors in code points based on static analysis rather than those that are focused on bytecode analysis which focuses on finding a series of specific vulnerabilities.

One of the most common problems is the stability of the tools over time. Many of them are created with a mere academic purpose, becoming outdated due to the evolution of libraries, frameworks, Solidity and compiler versions, causing them to stop being useful in the future. In addition, many of them are focused for an advanced user/expert in Solidity and Ethereum for several reasons:

- The analyst must know Solidity and also the different modifications that have been made over time. This is so because each tool works with a different version of the code and therefore, one tool may contemplate an error that another one does not.

- You must have full knowledge of the flaws that can occur when programming. The analyst must know the past of the Ethereum to know why the search of that type of vulnerabilities, not only it is enough to have learned Solidity recently.

- Use of libraries specifically designed for the purposes of these tools. During the installation and use of these tools, a lot of time has been lost in the installation process. Making a clean installation is expensive since one often depends on specific versions and in case a tool has a dependency on the same tool it can lead to failures. This makes the use of the containers provided by the developers almost mandatory.

*2.1. Related work: existing tools for analyzing vulnerabilities in Smart Contracts*

A detailed explanation of tools used in ESAF, can be read in [17] [25] . As an overview of all existing tools, Table 2 shows most of the existing tools classified according to two criteria: smart contract language and analysis method.

Being more specific in each tool, we have encountered the following difficulties in its installation or handling:

- **Solgraph** [26]: It should be installed with the following text to avoid errors:

    ```
    --unsafe-perm=true  --allow-root
    ```

- **Smartcheck** [27] : It did present no difficulty.

- **Solmet** [28]: It has a clear dependence with Java version number 8, in case we have any other version installed inside the computer where we are running the tool we will have to install that version. To be able to select a different version if we have several installed in the device it is recommended to use the following command.

    ```
    $sudo update-alternatives --config java
    ```

- **Oyente** [29]: It presents quite a few complications if you want to make a clean installation from scratch. You must use specifically the version of *Z3 Prover* that the creator asks of us. In addition, we must be careful with the version of the SOC since future versions may cause the tool to stop working.

- **Osiris** [30]: It works the same as Listener since it is based on it.

- **EthIr** [31]: This tool has a similar dependency on the 'Z3 Proverbs' library. The problem is that the previous tools coexist with EthIr since the versions of the library they support are different. This is one of the main reasons for choosing the containers to develop the tool. Also, the last tested version of the compiler was 0.4.25, being the last one 0.6.

- **Vandal** [32]: Given the requirements of Vandal for the installation of the tool it does not present a great difficulty except for its own use. Vandal is a tool for advanced users and is not recommended for new developers.

- **Mythril** [33]: Among the tools that have given problems at the time of installation natively Mythril has been one of them.

    On the other hand, the container they provide has certain limitations for its use. The tool fails due to the compiler version (0.5.1). For older contracts (compiler versions like 0.4.24) it is necessary to enter the contract version so that the tool can download the corresponding SOC

In addition, in order to be able to run the tool from the Python entrypoint set by the developers has been removed. This variable generates that the introduced command stops being of type bash to be only introduced commands that Mythril understands.

- **Securify** [34] : Its first version presented quite a few problems. Since we are currently using version 2 which is more stable, we have not found it to be a great difficulty if you follow the instructions.

- **Slither** [35]: Just like Mythril, with Slither the 'Text File' has been modified to remove certain restrictions. Among them, a user that created the container by default and prevented the use of the volumes for reading contracts has been removed.

- **Manticore** [36]: You have to specify all the versions of the compiler you work with. If you do not specify the one that has the smart contract, it will give an error and you will not be able to analyze it. It also has limitations with the number of versions since it does not work with all of them.

- **Madmax** [37]: It did present no difficulty.

- **Contract Larva** [38]: Although it is not included in the tool, it has been considered interesting to include it since it is a very useful tool. The main problem it presents is the installation of Haskell since the default installer installs a version that is not the correct one. It is recommended to use it:

```
$curl https://get-ghcup.haskell.org -sSf | sh
```

Table 3 summarizes the process of installing and launching all the tools. Column headers are selected thinking about the real application of each tool, outside from the theoretical and hypothetical world. The range selected for categories ease of installation and usefulness are 1 to 5. Meaning: 1 hard installation/useless to 5 easy installation/full utility respectively.

## 3. ESAF: Ethereum Security Analysis Framework

Following a comprehensive collection of most of the vulnerability analysis tools in smart contracts, with this tool we intend to not only unifies the analysis capabilities of all the tools but also provides a simple analysis interface that enhances the capabilities of each tool separately and can be used as a tool for monitoring and analyzing the persistent security of a set of smart contracts or as a tool for analyzing smart contracts individually.

ESAF's design principles are outlined below.

1. Use of containers to eliminate dependencies and encourage isolation.

2. It is scalable and support modifications.

3. Use as many tools as possible, the more the better.

4. Use of recent technologies.

5. Agility due to the volume of analysis.

### 3.1. Environment and Technologies

In this section, a small description is given of the environment and technologies that, due to their suitability and familiarity on the part of the authors, have been chosen for the development of the framework. The main problem was that each tool worked practically with its own dependencies and some of them concluded with others, leading to the use of containers after the analysis of these dependencies.

1. **Python:** Given the large number of tools that use Python [39] in both versions 2 and 3, it was decided to approach development through this language. Python has provided a convenient way to make a robust script by making use of the libraries it provides, in particular:

    - Docker SDK: It's the library that has needed the most research. The documentation and use is not very extended, surely caused by the short life of this and the changes that its creators have been doing. Frequently,

when it comes to solving execution errors, the solutions found on the Internet have not been useful due to changes in functions, input parameters, etc.Removing those kinds of exceptions, Docker SDK provides a very reliable way to use and handle dockers, providing most of the console commands in the form of function calls that can be found in the official documentation.

- MongoDB library: This is the official MongoDB library. It provides all the functions that can be found in the mongo CLI, making it very easy to implement connections to DB.

2. **Docker:** The technology chosen for the use of containers has been Docker [40]. Docker provides us with a standard software that packages or isolates the code and all its dependencies allowing the applications to run in a faster way and also gives us the possibility to distribute that image of the container regardless of the architecture and the operating system of the host.

3. **MongoDB:** In view of the storage of raw information, it was considered to use a non-relational database [41] for the storage of the contracts that were downloaded from the Ethereum node. Therefore, following the line taken, another database was created in MongoDB for the storage of the different analyses that were being used.

*3.2. Implementation*

The developed structure of the code has been based on the modulation and differentiation of the different files according to their functionality. In addition, it has sought to implement some design patterns and maintain some standards such as the injection of dependencies and object-oriented programming. Pseudocode of all the modules described in Table 4 can be found in https://github.com/alopezvivar/ESAF-PSEUDOCODE).

## 4. The framework in operation

ESAF is a framework that works in several phases, from obtaining the source code of the smart contracts, to the presentation of the results and various statistics of the results of the analysis made. This section describes these phases.

*4.1. Ethereum Node Deployment*

To obtain the source code of a smart contract already published in the Ethereum blockchain there are basically two ways. Either use the API of a blockchain explorer service such as Etherscan.io, Etherchain, etc. or deploy our own Ethereum node, synchronize it and extract the source code from the smart contract. The advantage that APIS have is the ease of use, but they have limitations of use so for the development of ESAF we chose the second option.

There are several clients of Ethereum as you can see in the Table 5 After trying go-ethereum with bad results (the node was not fully synchronized), we tried Parity and were luckier. However the process of synchronizing an Ethereum node is a process that can take several days (depending on hardware resources). Because of the way the information is stored in the blockchain, many random data are written to disk, so using an SSD speeds up the process compared to a conventional HDD. Using a large amount of RAM also helps speed up the node synchronization process.

In Figure 1 you can see the parameters used to synchronize our Ethereum node. The node runs in a virtual machine with Ubuntu Server 18.04 and 2TB of HDD and 32 GB of RAM The approximate time to fully synchronize the node was one week. It is interesting to mention that most modern Ethereum clients are able to synchronize the node faster by using a feature of the Ethereum protocol that allows downloading snapshots of the state of the blockchain and completing the synchronization from the genesis block to the current background block later. In parity this option is known as warp sync, but it has the disadvantage that although the node synchronizes faster, there are gaps (blocks) in the middle of the string that are not completely synchronized and if we want to extract the source code from a contract that is in one of those blocks we cannot until the block is not really synchronised in our node.

Figure 1: Parity config file

*4.1.1. Smart Contract Source Code Extraction*

Before we go on talking about the process of extracting source code from contracts (and why it's a slow process), it's important to stop a bit to explain what information and how it's stored in the Ethereum blockchain (and what information is stored "outside the chain").

In Figure 2 you can see how in Ethereum the information is stored in different Merkle Patricia Tries [42]. In Ethereum there are two types of data. The permanent data (for example transactions) and the ephemeral data (for example the balance of a particular account). Both types of information are not stored in the same place. Below is a brief description of each of the tries Ethereum uses to store information.

1. **State Trie:** This trie is the main Ethereum where the global status of the whole blockchain is stored. It is a trie that is constantly being updated and contains a key-value for each account that exists in the Ethereum network. The key is a unique identifier of 160bits and the value stores (in a coded form using RLP): a nonce, the account balance, a pointer to the root of the storage trie and a codehash (hash with the source code of the bytecode of the contract in the case of contract type accounts).

2. **Storage trie:** Each Ethereum account has its own trie to store all data associated with that account.

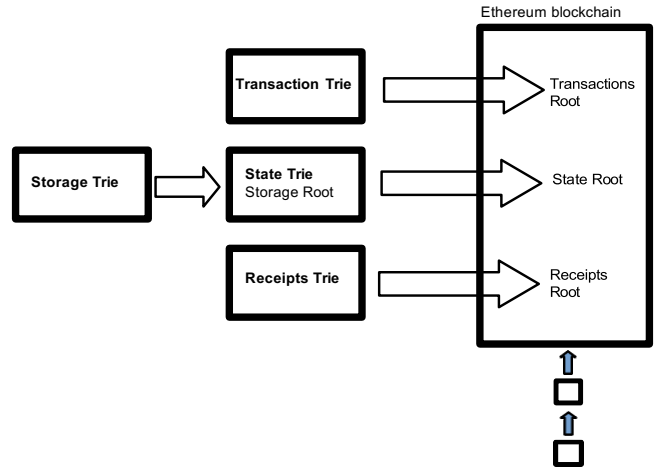

Figure 2: Ethereum blockchain data structure

3. **Transaction Trie:** Each block of Ethereum contains its own transaction trie. A block has many transactions. The order of the transactions in a block is of course decided by the miner who assembles the block. The route to a particular transaction in the transaction trie is via (the RLP encoding of) the index of where the transaction is located in the block. Mined blocks are never updated; the position of the transaction in a block is not changed. This makes it possible to return to the original path over and over again to recover the original result once a transaction is located in a block's transaction trie.

4. **Receipts Trie:** The transaction receipt trie has all the transaction receipts for the transactions included in a block. The hash of the root node of the transaction receipts trie is included in the block header (in the receiptsRoot field) There is one transaction receipts trie per block.

For the task of exporting the source code of smart contracts from the Ethereum blockchain has been made use of the tool *ethereum-etl* [43] In order to be able to extract the source code from the contracts, it is necessary to first extract the addresses from the contracts. Now, to get the addresses of the contracts previously you have to extract the hashes of the transactions but to have the hashes of the transactions previously you have to export the blocks. As you can see, the process of extracting the source code of the contracts from the blockchain is slow because there is a lot of indirection and it requires processing time. In *ethereum-etl*, the command *export_contracts* allows to perform the steps mentioned above in a sequential way. In Figure 3 and Figure 4 you can see part of the extraction process.

Figure 3: Extracting blocks 7700000 to 7799999

Figure 4: Extracting token transfers from blocks 7700000 to 7799999

*4.1.2. Processing Smart Contracts Source Code*

The ethereum-etl tool extracts the contracts in csv files and organized by blocks and directories. Once these files were generated, a tool was written that processes the directories and extracts from the csv files the source code of the contracts and inserts them in a non relational database. This process can be seen in the Figure 5 The etherscan api is also used in this phase to try to obtain the source code of the contract written in Solidity since in the blockchain it is only stored in bytecode.

Figure 5: Processing smart contract source code and inserting in MongoDB

*4.1.3. Smart Contracts Vulnerability Analysis*

It is in this phase that the contracts with the developed meta-tool are analysed. Figure 6 shows the process of preparing for the analysis up to the moment before starting to iterate on the collection of selected contracts. In it you can see the selection of the ranges to be analysed (or the whole database), the creation of the input files (taking the information from mongoDB). Figure 7 shows the following main functionality of the application, iterating through the cursor returned by MongoDB we extract the contract where the Solidity code and the bytecode are located.

The process is as follows:

1. Smart contract code is added to an input file and stored in the inputs folder.

2. Each docker will have a shared volume, common to all dockers, where you can read these files.

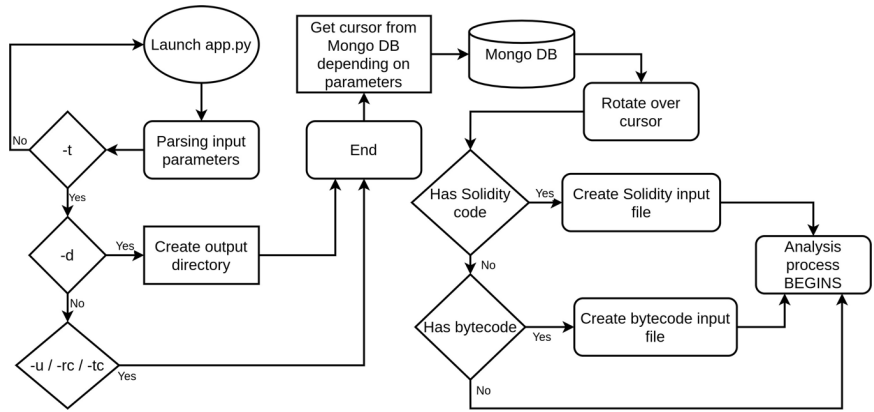

Figure 6: ESAF pre analysis flow chart

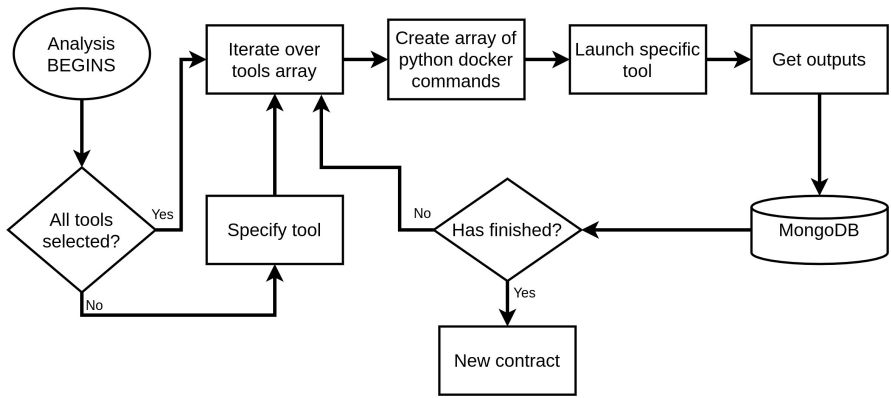

Figure 7: ESAF analysis flow chart

3. Later, an array of commands will be create and passed to the docker depending on the tool.

4. Finally, the results of each tool will be collected and inserted in mongoDB.

5. Until all contracts and tools are iterated, the execution will not it will end.

Finally, the Figure 8 shows a sequence diagram of the application.

Tools that run Solidity are much more likely to fail due to the clear dependency on the Ethereum compiler commonly called Solc. The problem is to find code that does not have a version of the compiler installed in the specific docker of the tool, to avoid failures we have chosen tools that support Solidity but do not have a compiler dependency. On the other hand, the tools that support bytecode offer a very good performance without giving failures.It has been observed how they report tool-specific bugs.In the analysis of 1000 elements there has not been any kind of error, reporting failures in each one of them. Finally in Listing 1 you can see the analysis output of a smart contract using our tool.

Table 1: Types of vulnerabilities in smart contracts

| Vulnerability | Description |
|---|---|
| Reentrancy | This is a vulnerability well known for its impact. The programmer may think that a non-recursive function cannot be re-called while it is running, but this is not always the case, as it could be the case that within the function an empty malicious contract is called that only contains a function of *fallback* that calls back the function it comes from. |
| Exception disorder | The treatment of exceptions in Solidity has particularities depending on whether the exception occurred in a call to a method directly or using the CALL primitive, which can cause vulnerabilities if a malicious user causes exceptions to alter the execution flow. |
| Calls to the unknown | Some methods of Solidity when invoked can call other methods of contracts that in principle are unknown and with potentially dangerous side effects. |
| Type conversion | Although the Solidity compiler can detect errors with types, for example if a function waits for an integer and is called by passing it a string, in the case of contract definitions or functions with a certain structure, in the case of calling a function in a contract, if the programmer makes a mistake and calls by mistake another contract but it contains a function with the same structure expected by the compiler, the function will be executed and if the function does not exist, the function of *fallback* will be called. In any case, no exception will be launched. |
| Secrets | Solidity allows you to define the visibility of fields in contracts as public or private. This can be useful if you need to hide certain information between contract calls. Unfortunately this system is not effective as changes in private fields have to be sent to mining nodes to be put into blockchain, which is public. |
| Unpredictable state | All smart contracts contain a status based on the values of their fields and their balance sheet. However, there is no guaranteed that the state of a contract when we performed a transaction is the identical to the state when that transaction is pulled out and placed in the blockchain. In other words, it could happen that previous to processing our transaction, other transactions have already changed the state of the target contract and in addition to being fast, it does not guarantee us anything since the miners can mine the transactions in the order they wish. There is a further problem caused by the nature of blockchain, which is that a chain fork could occur if two miners continue to mine a valid block at the same time. This would make some miners try to put their block on one of the two chains and the others on the other. At any moment the shorter chain would be dismissed, losing the transactions contained in it and changing the state of the contracts to an indeterminate state. |
| Random numbers | The execution of the Ethereum virtual machine code is deterministic. This means that the code executed with the same inputs must produce the same output in all the nodes that execute it. This presents a problem when generating random numbers. To simulate randomness, many contracts use a random number generator initialised with the same seed for all miners. |
| Time restrictions | Many applications have time restrictions to operate. Usually these restrictions use texttimestamps. In the case of smart contracts, the programmer can get the *timestamp* of when the block was mined, which is shared by all transactions in the block. The trouble is that miners in the early versions of the protocol could choose the *timestamp* of the block they were going to arbitrarily mine, which could be used to carry out attacks. |
| Immutable bugs | This is not a vulnerability in itself, but the consequence of a blockchain property. All the source code of smart contracts, including those containing *bugs* are immutable once they are mined and added to the blockchain, although they can be blocked by calling a destructor function. |
| Loss of Ether | If the developer is wrong to enter the address to send *ether* and that address exists but it is an orphan address that belongs to no one that *ether* will be lost forever. |
| Stack size | Each time one contract calls another contract the associated call stack increases by one. The stack limit is 1024 and when the limit is reached an exception is launched. Until October 18, 2016 it was possible to take advantage of this to launch an attack where a malicious user increased the battery counter until almost exhausted and then called the victim's function which launched an exception when the battery limit was exhausted. If the victim did not take this into account and does not handle the exception correctly, the attack could be successful. The impact of this vulnerability caused Ethereum to be redesigned. |

Table 2: Summary of security tools specifications

| | | Oyente | Solgraph | MadMax | Manticore | SmartCheck | Mythril | ContractLarva | SolMet | Vandal | EthIR | MAIAN | Erays | Rattle | Osiris | Securify | Slither | Ethertrust |
|---|---|---|---|---|---|---|---|---|---|---|---|---|---|---|---|---|---|---|
| Level | Bytecode | ✓ | X | ✓ | ✓ | X | ✓ | X | X | ✓ | ✓ | ✓ | ✓ | ✓ | ✓ | ✓ | ✓ | ✓ |
| | Solidity | X | ✓ | X | X | ✓ | X | ✓ | ✓ | X | X | X | X | X | X | X | ✓ | X |
| Analysis | Dynamic analysis | X | X | X | X | X | X | ✓ | X | X | X | ✓ | X | X | X | X | X | X |
| | Static analysis | X | X | X | X | X | ✓ | X | X | X | X | X | X | X | X | X | ✓ | X |
| | Formal verification | ✓ | X | ✓ | ✓ | X | ✓ | X | X | ✓ | ✓ | ✓ | X | X | ✓ | ✓ | X | ✓ |

Table 3: Summary of the installation of the security tools

| Tool | Ease of Installation | Usefulness | Stays up to date | Dependencies |
|---|---|---|---|---|
| Oyente | 1/5 | 3/5 | No | python2 (pip2), z3 prover, web3 (pip3), solc, Go-ethereum |
| Solgraph | 4/5 | 5/5 | No | nodejs, npm, graphviz |
| Madmax | 4/5 | 5/5 | No | python3 |
| Manticore | 5/5 | 5/5 | Yes | python3 (+3.6v), pip3, solc |
| SmartCheck | 5/5 | 5/5 | Yes | npm |
| Mythril | 5/5 | 5/5 | Yes | npm, python3, pip3 |
| ContractLarva | 3/3 | 3/3 | Yes | Haskell (ghc) |
| SolMet | 5/5 | 5/5 | No | java, maven, CSV reader |
| Vandal | 4/4 | 4/4 | No | python3 |
| EthIR | 1/5 | 3/5 | Yes | python2 (pip2), z3 prover, web3 (pip3), solc, Go-ethereum |
| MAIAN | 1/5 | 1/5 | No | solc, z3, python3, web3 |
| Erays | 5/5 | 3/5 | No | graphviz, python |
| Rattle | 5/5 | 5/5 | No | graphviz, solc, python3 |
| Osiris | 2/5 | 5/5 | No | python2 (pip2), z3 prover, web3 (pip3), solc, Go-ethereum |
| Securify | 5/5 | 5/5 | No | souffle´, java 8, solc |
| Slither | 5/5 | 5/5 | Yes | solc, python3 |
| Ethertrust | 1/5 | 2/5 | No | z3 prover, python, maven |

Table 4: Scripts and description used by the tool

| File | Description |
|---|---|
| app.py | Main module of the application, test launch and argument parsing. |
| app-engine.py | Generates an array of executions based on an input and a tool/s. |
| dependencies-builder.py | Initially, the idea was to create an output file for each execution made on an smart contract. The module would be in charge of joining the creation of all the files on which each tool depended. Finally, having hosted the results of the database its main function is to create the input files in bytecode and Solidity code. |
| md-contracts.py | Module responsible for instance the connection to the database, is responsible for returning the cursor is according to the search parameters, database updates, delete items, etc. |
| contract-class.py | Template used by the application for instance of a contract object. It has "getters" and "setters" methods. |
| constants.py | The different program constants, paths to directories, available tools, etc. are hosted. |
| exceptions.py | File with the tool's own actions. |
| stack-trace.log | File where the logs (info, warning, exception) of the application are stored. |

Table 5: Ethereum clients list

| Client | Language | Developers | Latest release |
|---|---|---|---|
| go-ethereum | Go | Ethereum Foundation | go-ethereum-v1.4.18 |
| Parity | Rust | Ethcore | Parity-v1.4.0 |
| cpp-ethereum | C++ | Ethereum Foundation | cpp-ethereum-v1.3.0 |
| pyethapp | Python | Ethereum Foundation | pyethapp-v.1.5.0 |
| ethereumjs-lib | Javascript | Ethereum Foundation | ethereumjs-lib-v3.0.0 |
| Ethereum(J) | Java | ether.camp | ethereumJ-v1.3.1 |
| ruby-ethereum | Ruby | Jan Xie | ruby-ethereum-v0.9.6 |
| ethereumH | Haskell | BLockApps | no release yet |

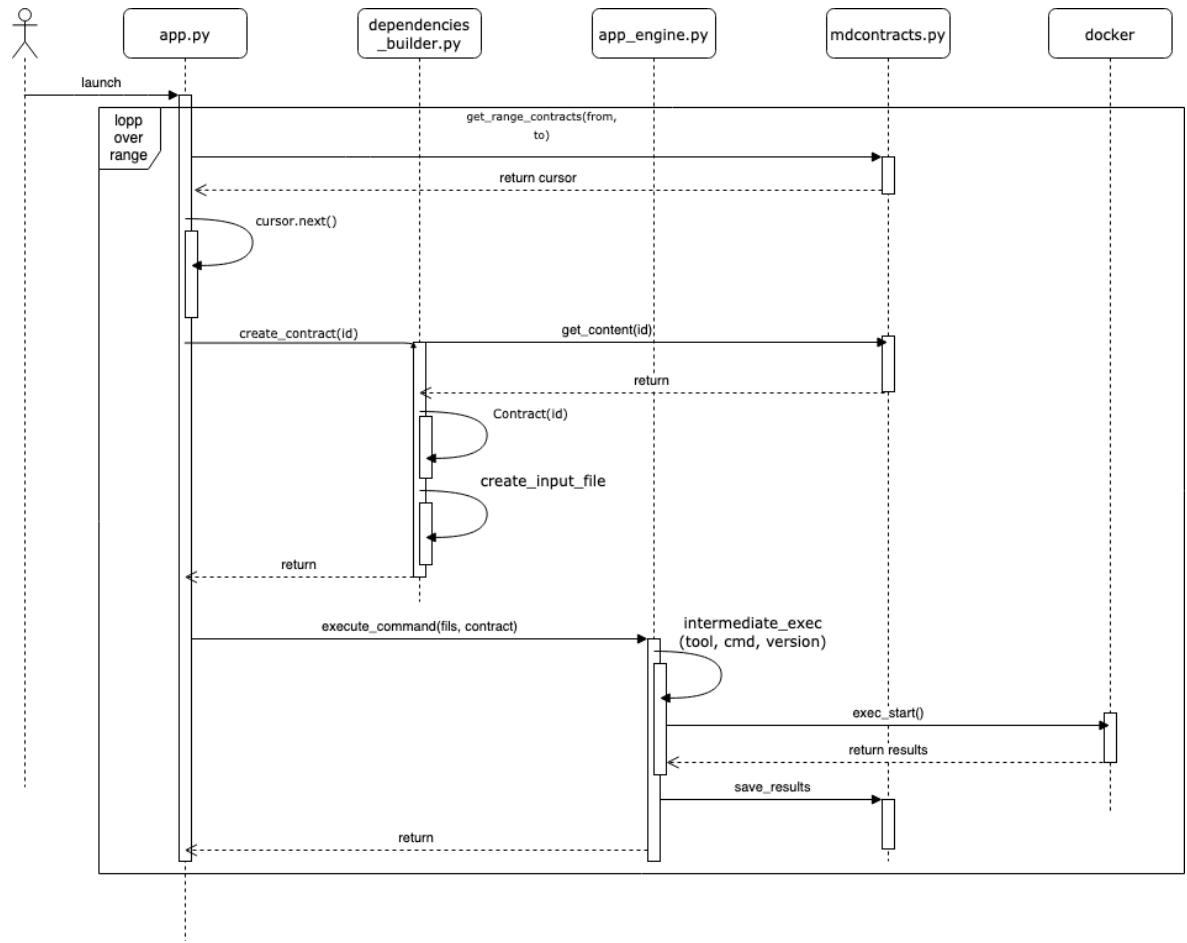

Figure 8: Sequence diagram of the analysis tool

Listing 1: Example of results of a vulnerability analysis using ESAF

```
{
  "_id":"ObjectId(" "5e9c6c97574ce3bef34d9f15" ")",
  "contract_id": 200,
  "address" : "0x2194b1734ee0f67440884da49952a45b34ba832d",
  "solgraph":{
     "output":"strict digraph {\n   mul [color=gray]\n   div [color=gray]\n   sub [color=gray]\n   add [color=gray]\n
          ↪    totalSupply [color=yellow]\n   ... n",
     "time_elapsed":0.021505345000008447
  },
  "smartcheck":{
     "output":"input_contract_200.sol\n\njar:file:/usr/local/lib/node_modules/@smartdec/smartcheck/jdeploy-bundle
          ↪ /smartcheck-2.0-jar-with-dependencies.jar!/solidity-rules.xml\nruleId: \nSOLIDITY_ERC20_APPROVE\
          ↪ npatternId: \naf782c\nseverity: \n2\n\nline: \n90\n\ncolumn: \n4\ncontent: \nfunctionapprove(
          ↪ address_spender,uint256_value)publicreturns(bool){allowed[msg.sender][_spender]=_value;
          ↪ emitApproval(msg.sender, _spender, _value ...",
     "time_elapsed":0.05430566500000111
  },
  "solmet":{
     "output":"Parsing /tmp/inputs/input_contract_200.sol\n\nSolidityFile;ETHAddress;ContractName;Type;SLOC;LLOC;
          ↪ CLOC;NF;WMC;NL;NLE;NUMPAR;NOS;DIT;NOA;NOD;CBO;NA;NOI;A...",
     "time_elapsed":0.02934941899999899
  },
  "oyente":{
     "output":"Contract extension doesn't allow this analysis",
     "time_elapsed":"0"
  },
  "osiris":{
     "output":"Contract extension doesn't allow this analysis",
     "time_elapsed":"0"
  },
  "ethir":{
     "output":"Contract extension doesn't allow this analysis",
     "time_elapsed":"0"
  },
  "vandal":{
     "output":"Contract extension doesn't allow this analysis",
     "time_elapsed":"0"
  },
  "mythril":{
     "output":"The analysis was completed successfully. No issues were detected.\n\n\n",
     "time_elapsed":0.026984572999992906
  },
  "securify":{
     "output":"",
     "time_elapsed":0
  },
  "slither":{
     "output":"INFO:Detectors:\\x1b[92m\ncustomIcoToken.createTokens()   (../../../tmp/inputs/input_contract_200.
          ↪ sol#127-142) compares to a boolean constant:\n\t-require(bool)(isFinalized == false) (../../../tmp
          ↪ /inputs/input_contract_200.sol#128)\ncustomIcoToken.finaliz...",
     "time_elapsed":0.03750889799999868
  },
  "manticore":{
     "output":"2020-04-19 15:34:44,416: [6] m.main:\\x1b[34mINFO:\\x1b[0m Registered plugins:
          ↪ DetectIntegerOverflow, DetectExternalCallAndLeak, DetectEnvInstruction, DetectUninitializedStorage
          ↪ , DetectUninitializedMemory, DetectReentrancySimple, DetectReentrancyAdvanced,
          ↪ DetectManipulableBalance,DetectUnusedRetVal, DetectSuicidal, DetectInvalid, DetectDelegatecall\n\
          ↪ n2020-04-19 15:34:44,417:[6] m.main:\\x 1b[34mINFO:\\x 1b[0mBeginning analysis\n\n 2020-04-19 15:34
          ↪ :44,444: [6] m.e.manticore:\\x1b[34mINFO:\\x1b[0m Starting symbolic create ...",
```

```
    "time_elapsed":0.0227155660001645
  },
  "madmax":{
    "output":"Contract extension doesn't allow this analysis",
    "time_elapsed":"0"
  }
}
```

*4.2. User Interface*

The developed tool is handled mainly by command interface although a very basic graphic interface has also been added, which will be improved in the future with new functionalities. In Figure 9 you can see the original model. First the user can select which type of scan he wants whether to scan local files (Figure 10) or a particular address (Figure 11). The user then selects those tools that they want to perform a scan. A series of buttons will be available for quick dialing of these values, being free to mark or unmark the ones you want. The tool supports both .hex and .sol format contracts.

In case the user selects an analysis by means of an address, those elements that do not depend on this type of analysis will be removed, such as the selection of contracts by the employer and the list of these selected files, and an input field will be added where the user can enter the address to be analysed. It must be taken into account that if the address added is not specific to an smart contract, it will not produce any result since this type of address does not have any type of code, they are user accounts.

## 5. Conclusions and Future Work

With this work the authors have presented a framework that makes it easier for developers of smart contracts to analyze vulnerabilities in their contracts by combining the power of many static/dynamic code analysis tools already published and tested by the community and allowing to add or remove tools that are appearing or becoming obsolete without having to worry about installation problems or requirements of each tool separately.

Another use of ESAF that we think is interesting is as a persistent "pentesting" tool, allowing us to monitor smart contracts of our interest as well as their interactions with other contracts for anomalies that may be detected as vulnerabilities in the future or to monitor smart contracts that although at the time of their deployment in the block chain they did not contain vulnerabilities, later on thanks to improvements in analysis techniques they could turn out to be insecure.

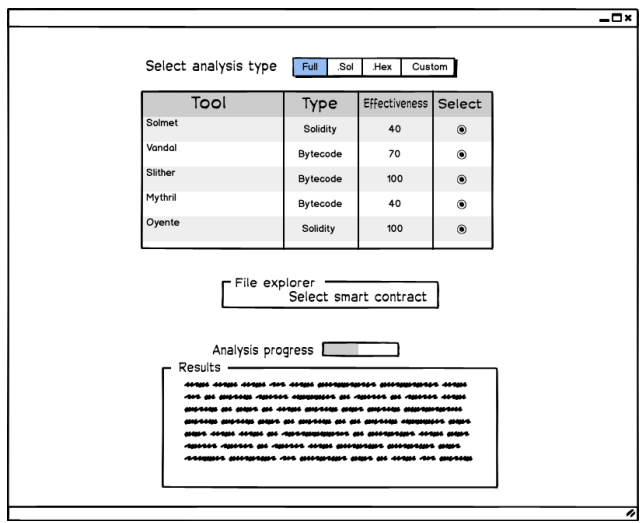

Figure 9: Preliminary interface

The multitude of existing security tools have advantages over each other, but generally, due to the continuous modifications of the Solidity compiler there is a great coupling to the version with which the tool is developed. The more modifications made to the primitive code (since it has not yet reached version 1.0) the more problems there will be with the tools that depend on it, compromising security.

In our first ESAF tests the tools against high load jobs (ten thousand analysis at a time) have presented an optimal performance in spite of not having the parallelization implemented. If parallelization is added and the number of containers and resources is increased, the speed of analysis will increase enormously. The scans thrown during the tests were satisfactory, vulnerabilities were found in most of the contracts. This is because the first part of the downloaded smart contract database contained older versions of the compiler. As for the effectiveness of mass scans, the network has certain limitations. A large number of contracts cannot be easily downloaded. The main reason is the high cost of node synchronization with the network and the high demand for information.

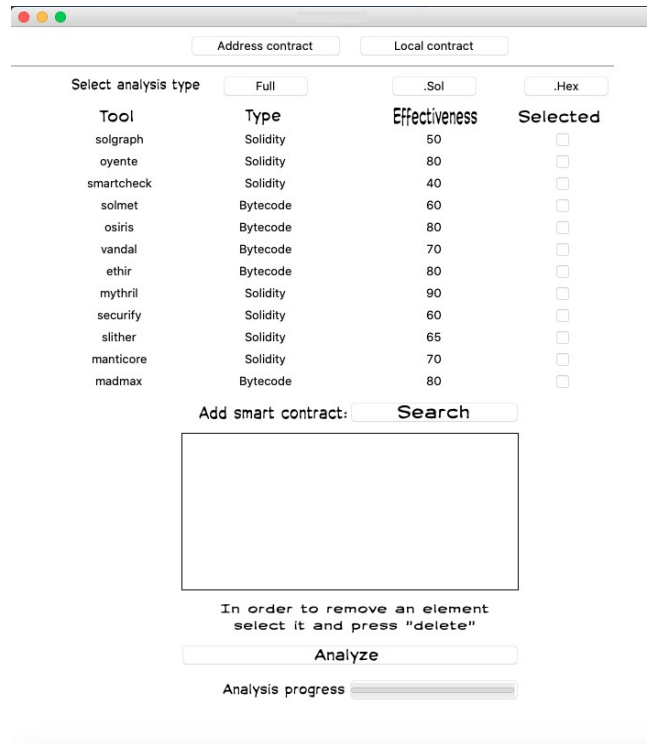

Figure 10: Selecting the local analysis option

## 5.1. Future Work

The meta-tool developed for this work is fully functional, although like all software, it can be improved and its functionality extended. The main line of future work is to add a module to the meta-tool that allows obtaining statistics related to smart contracts such as:

- Percentage of contracts that executed each tool

- Percentage of contracts with arithmetic related vulnerabilities

- Percentage of contracts with transaction related vulnerabilities

- Percentage of contracts with access and visibility related vulnerabilities

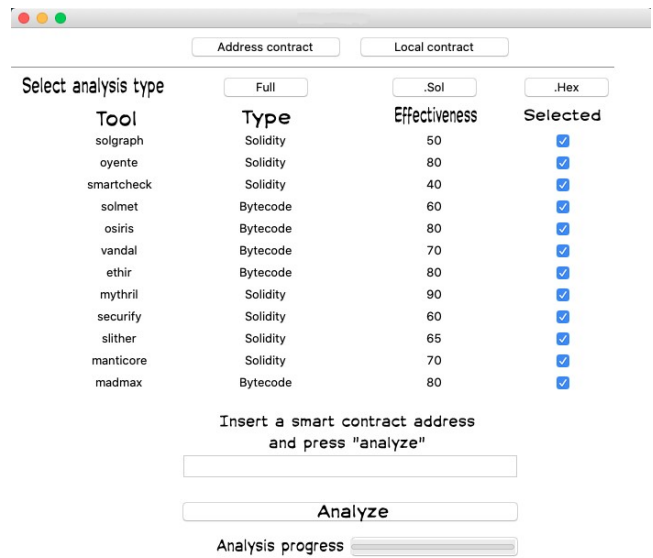

Figure 11: Address analysis

- Percentage of contracts using selfdestruct method

- Percentage of contracts implementing cipher methods to store variables

- Percentage of contracts with outdated compiler version

- Percentage of contracts using delegatecall methods

- Percentage of contracts using Safe Math library

- Percentage of contracts using external libraries

- Relationship between contract size and vulnerabilities found

- Relationship between compiler version and vulnerabilities found

- Which different metrics each tool found

- For each vulnerable contract, compare tools' results

- Compare tools by bytecode and solidity

- Overview of contracts vulnerabilities detected over time

On the other hand it is very likely that new tools for vulnerability analysis in smart contracts will continue to appear and could be added to the current ones, in addition to other tools that will be updated or cease to exist. The graphical interface, right now in an initial version could also be extended with options for analysis by time ranges for example.

Finally, there is an article in progress for the development of a vulnerability analysis tool in Ethereum using machine learning algorithms where the meta-tool presented here will be used in the phase of tagging vulnerable contracts, in order to cover a greater number of vulnerabilities.

**Acknowledgements**

This project has received funding from the European Unions Horizon 2020 research and innovation programme under grant agreement No 700326. Website: http://ramses2020.eu